\documentclass[11pt]{article}
\usepackage{graphicx}
\usepackage{amssymb}
\usepackage{amscd}
\usepackage{mathrsfs}
\usepackage{longtable,lscape}
\usepackage{amsthm}
\usepackage{amsfonts}
\usepackage{amsmath}
\usepackage{bbm}
\usepackage{float}
\usepackage{url}
\usepackage{csquotes}
\usepackage{wrapfig}

\usepackage{caption}
\usepackage{lineno}
\usepackage[title]{appendix}
\begin{document}
\title{\bf Violation of the Bell-CHSH inequality and marginal laws in a single-entity Bell-test experiment}
\author{Diederik Aerts and Massimiliano Sassoli de Bianchi\vspace{0.5 cm} \\ 
\normalsize\itshape
Center Leo Apostel for Interdisciplinary Studies, \\ \itshape Brussels Free University, 1050 Brussels, Belgium\vspace{0.5 cm} \\ 
\normalsize
E-Mails: \url{diraerts@vub.ac.be}, \ \url{msassoli@vub.ac.be}
}
\date{}
\maketitle
\begin{abstract} 
\noindent We describe a simple experimental setting where joint measurements performed on a single (classical or quantum) entity can violate both the Bell-CHSH inequality and the marginal laws (also called no-signaling conditions). Once emitted by a source, the entity propagates within the space of Alice's and Bob's detection screens, with the measurements' outcomes corresponding to the entity being absorbed or not absorbed in a given time interval. The violation of the marginal laws results from the fact that the choice of the screen on the side of Alice affects the detection probability on the side of Bob, and vice versa, and we show that for certain screen choices the Bell-CHSH inequality can be violated up to its mathematical maximum. Our analysis provides a clarification of the mechanisms that could be at play when the Bell-CHSH inequality and marginal laws are violated in entangled bipartite systems, which would not primarily depend on the presence of a bipartite structure but on the fact that the latter can manifest as an undivided whole.
\end{abstract} 
\vspace{0.5cm}

\section{Introduction}

\begin{figure}[!ht]
\centering \includegraphics[width =12cm]{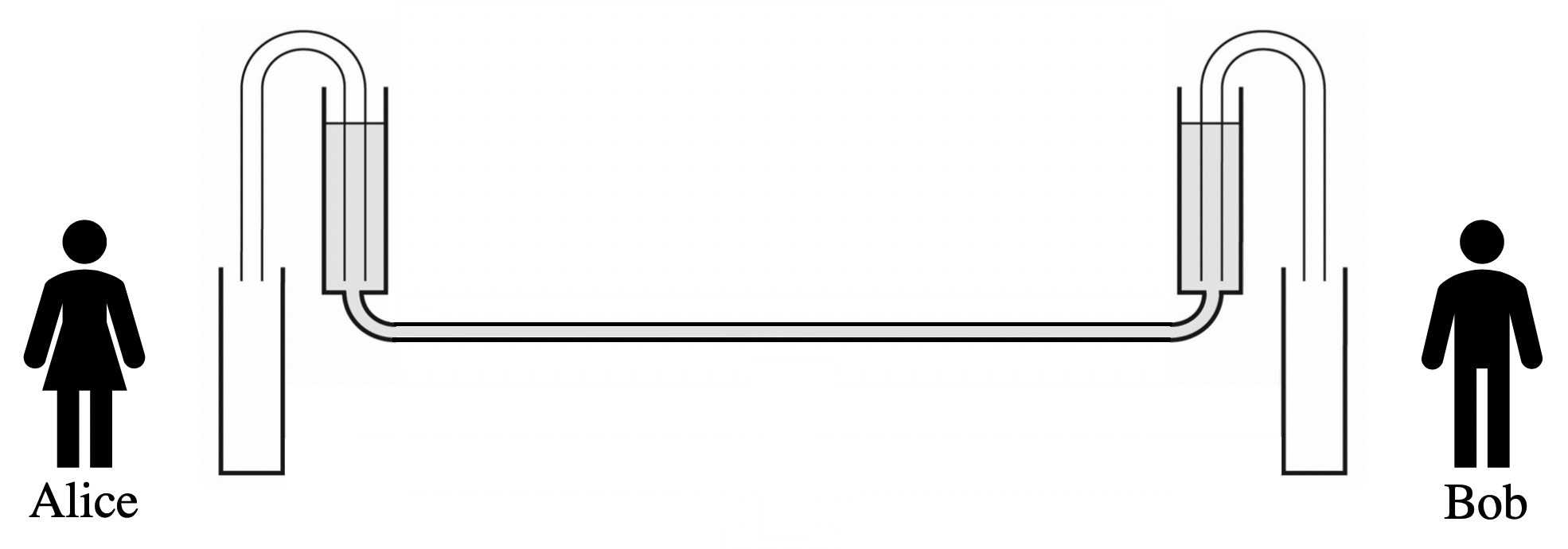} 
\caption{A schematic representation of the vessels of water experiment.
\label{Figure1}} 
\end{figure}

In the eighties of last century, one of us imagined a way to perform coincidence measurements on a macroscopic system formed by two vessels connected through a tube and containing a certain amount of transparent water 
\cite{Aerts1982,Aerts1983}, obtaining in this way an unexpected maximal violation of the Clauser, Horne, Shimony and Holt (CHSH) version of Bell's inequality \cite{clauser1969,Ballentine1987a,Alford2016} (see Figure~\ref{Figure1}). More precisely, in this experiment Alice and Bob measurements $A$ and $B$ consist in pouring out the water from their respective vessels using a siphon, into reference vessels. The positive outcome is when they can collect more than half of the total quantity of water (let us denote these positive outcomes $A_1$ and $B_1$, respectively) and of course they cannot jointly obtain positive or negative outcomes, i.e., their outcomes are perfectly anti-correlated. Alice and Bob additional measurements, $A'$ and $B'$, consist in simply checking the transparency of the water, and since the water used in the experiment is actually transparent, they will always obtain positive outcomes (which we will denote $A'_1$ and $B'_1$, respectively). So, if we denote $A_2$ and $B_2$ the negative outcomes of the $A$ and $B$ measurements, respectively, and $A'_2$ and $B'_2$ the negative outcomes of the $A'$ and $B'$ measurements, respectively, we have the probabilities: 
\begin{eqnarray}
&&{\cal P}(A_1,B_1)=0,\quad {\cal P}(A_2,B_2) = 0,\quad {\cal P}(A_1,B_2)={1\over 2},\quad {\cal P}(A_2,B_1)={1\over 2},\nonumber\\
&&{\cal P}(A_1,B'_1)=1,\quad {\cal P}(A_1,B'_2) = 0,\quad {\cal P}(A_2,B'_1)=0,\quad {\cal P}(A_2,B'_2)=0,\nonumber\\
&&{\cal P}(A'_1,B_1)=1,\quad {\cal P}(A'_1,B_2) = 0,\quad {\cal P}(A'_2,B_1)=0,\quad {\cal P}(A'_2,B_2)=0,\nonumber\\
&&{\cal P}(A'_1,B'_1)=1,\quad {\cal P}(A'_1,B'_2) = 0,\quad {\cal P}(A'_2,B'_1)=0,\quad {\cal P}(A'_2,B'_2)=0.
\label{prob-vessels}
\end{eqnarray}
It follows that for the expectation values (also called correlation functions) we obtain: 
\begin{eqnarray}
&&E(A,B)={\cal P}(A_1,B_1) + {\cal P}(A_2,B_2) - {\cal P}(A_1,B_2)- {\cal P}(A_2,B_1)= -1\nonumber\\
&&E(A,B')={\cal P}(A_1,B'_1) + {\cal P}(A_2,B'_2) - {\cal P}(A_1,B'_2)- {\cal P}(A_2,B'_1)= 1\nonumber\\
&&E(A',B)={\cal P}(A'_1,B_1) + {\cal P}(A'_2,B_2) - {\cal P}(A'_1,B_2)- {\cal P}(A'_2,B_1)=1\nonumber\\
&&E(A',B')={\cal P}(A'_1,B'_1) + {\cal P}(A'_2,B'_2) - {\cal P}(A'_1,B'_2)- {\cal P}(A'_2,B'_1)= 1.
\label{expectations}
\end{eqnarray}
Hence, choosing one of the possible sign combinations for the quantity in the Bell-CHSH inequality \cite{clauser1969} (which can be obtained by interchanging the roles of $A$ and $A'$ and/or the roles of $B$ and $B'$), we find: 
\begin{eqnarray}
{\rm CHSH} &\equiv& E(A',B')+E(A,B') + E(A',B) - E(A,B) \nonumber \\
&=&1+1+1-(-1)=4,
\label{CHSH}
\end{eqnarray}
which is the maximum possible violation. 

The above example shows that not only the violation is not just the prerogative of micro-systems, but also that the fundamental mechanism producing it could be that of the `creation of correlations' during the joint measurement processes, as in fact the very quantum formalism also suggests. Correlations that are created during the measurement processes were given the name of `correlations of the second kind', to distinguish them from the more conventional `correlations of the first kind', which are already actual prior to the execution of the measurements and cannot violate the Bell-CHSH inequality. 

Note that correlations of the second kind have escaped the attention of most of the scientists analyzing situations violating Bell's inequalities, as they constitute a sort of ``singularity'' with respect to the way the situation is usually looked at. Indeed, Bell already introduced a demarcation between correlations originating from a common cause (which cannot violate his inequalities, like in his famous example of Bertlmann's socks \cite{Bell1981}) and correlations not originating from a common cause (which in principle can violate his inequalities). This demarcation proposed by Bell is of course a meaningful one, but what was overlooked at the time is that there is also the possibility that the experiment itself could create, during its execution, the common cause which in turn creates the correlations. And since in typical Bell-test experiments four different coincidence measurements are considered, if each of them creates its own `common cause', strictly speaking there will be no `unique common cause' for all the observed correlations so that the Bell-CHSH inequality can in general be violated.\footnote{This explains in part why the idea of `correlations of the second kind' has not received so far enough attention and is usually not included in the set of possible explanations of quantum correlations, despite the fact that it is a very simple and powerful explanation. In other words, the lack of notoriety of this simple mechanism of `actualization of a common cause' should not give rise to the idea that it would conceal some kind of error. It is a perfectly valid physical mechanism, which allows to explain in a very effective way the appearance of quantum correlations as correlations having a contextual common cause.}

Models that are similar in their conception to the vessels of water model were analyzed in the years (see for example \cite{Aerts1991,AertsDurt1994,aabg2000,Sassoli2013a,Sassoli2013b,Sassoli2013c,Sassoli2014}) 
and were instrumental in giving life to an approach to quantum mechanics called the `hidden-measurements approach' (or `hidden-measurements interpretation'), where quantum measurements are described as weighted symmetry breaking processes that can select a different `measurement-interaction' at each run of the experiment. In other words, in the hidden-measurements approach the hidden variables are to be associated with the interactions between the measured entities and the measuring apparatuses and not with the states of the former, which is the reason why the no-go theorems do not apply in this case. In more recent years, the approach was given a very general mathematical formulation called the `extended Bloch representation (EBR) of quantum mechanics' \cite{AertsSassolideBianchi2014,AertsSassoli2016,AertsSassolideBianchi2017,AertsSassoli2019}, using a generalization of the Bloch 
sphere representation of states, valid for physical entities of arbitrary dimension, with the measurement-interactions being described in the formalism as elements of specific geometrical structures. Note however that these hidden measurement-interactions, which allows to explain the wave-function collapse as an objective physical process and to derive in a non-circular way the Born rule, should not be understood in the sense of `fundamental forces' mediated by some kind of bosonic entities, describable for instance by potential terms in a Lagrangian. One has instead to think about them as effects resulting from the complex non-spatial/non-local fluctuations/perturbations that are generated by the very act of bringing a measured system in contact with a measuring apparatus. 

The description of entangled systems in the ambit of the EBR has shown that the vessels of water model is more than just a clever metaphor, but 
truly a paradigmatic exemplification of what possibly goes on behind the scenes of a Bell-test experiment \cite{AertsSassoli2016}. Of course, a system formed by two vessels of water connected through a tube is very different from, say, two entangled electrons in a singlet state, as for the two vessels of water the non-local connection between the two sub-systems -- the tube -- is entirely visible under our eyes (see Figure~\ref{Figure1}), whereas for two entangled electrons nothing can be detected in the space between Alice's and Bob's positions, once the electrons have been emitted by the source and have propagated away from it. In other words, two electrons in a singlet state are connected in a non-spatial way, something that is reflected in the EBR formalism by the fact that the vector describing the `element of reality of the connection' has a greater dimensionality than the vectors describing the individual states of the sub-entities forming the bipartite system \cite{AertsSassoli2016}.

In other words, the two connected vessels of water form a macroscopically whole spatial system that can manifest a quantum-like behavior when measurements are executed that can break its wholeness and by doing so create correlations, which in turn will violate the Bell-CHSH inequality. The violation produced by two entangled electrons could very well be of the same kind, with the difference that the element of reality able to make the bipartite entity a whole interconnected system would not be a spatial one, i.e., something that can be represented in our three-dimensional spatial theater. In our view, explaining the non-spatiality of the quantum micro-entities is one of the challenges that quantum physics poses, forcing us to reconsider what the nature of a physical entity might be, and we refer to \cite{Aertsetal2018}, and the references cited therein, for a speculative proposal. What is also important to emphasize here is that a quantum-like behavior is 
the expression of a specific form of organization that can be evidenced at a structural-modeling level, hence it can also appear at the macro-level of our physical reality \cite{AertsEtal2015,AertsSassoli2018b}.

Having made these due premises, the purpose of this article is to present and analyze a very simple and fundamental example of a physical system subjected to two-outcome joint measurements that can violate the Bell-CHSH inequality. Remarkably, the system in question is just an elementary single entity, which moreover can indifferently be either classical or quantum. Note that there has been some interest in the literature regarding the possibility of violating Bell's inequalities by means of single-entity entangled states of the form:
\begin{equation}
|\psi\rangle = {1\over\sqrt{2}}(|0\rangle\otimes |1\rangle +|1\rangle\otimes |0\rangle),
\label{sup1}
\end{equation}
where $|0\rangle$ describes a zero-entity state and $|1\rangle$ a one-entity state, the violation being possible if no superselection rules forbid the local superposition of a zero-entity with a one-entity state; see for instance \cite{Enk2005,Klyachko2005,PawlowskiCzachor2006,Cunhaetal2007,Ashhabetal2007a,Ashhabetal2007b,Wasaketal2018}. Our model is also a single-entity one, in which joint measurements can create significant correlations, but not because it would be entangled with a vacuum (zero-entity) state. 

Note also that, as it was the case for the historical vessels of water model, and similar models, there is a violation not only of the Bell-CHSH inequality but also of the marginal laws, also called no-signaling conditions (see \cite{AertsEtAl2019} and references cited therein). This goes apparently against the theoretical predictions of the standard quantum formalism. However, it is important to observe that the Bell-test experiments conducted in the laboratories have so far 
shown numerous violations of the marginal laws \cite{AdenierKhrennikov2007,DeRaedt2012,DeRaedt2013,AdenierKhrennikov2016,Bednorz2017,Kupczynski2017}. The debate about the origin of these violations is still open and there is the tendency to believe that they are only due to experimental errors. But regardless of what the final verdict will be, it is important to observe that quantum mechanics does not forbid such violations, and it is also one of the objectives of this article to highlight this aspect, showing that coincidence measurements performed on single-entity states will systematically violate the marginal laws and that this does not imply that superluminal signaling would for this be possible.

\section{The double-screen model}
\label{double-screen-classical}

\begin{figure}[!ht]
\centering \includegraphics[width =12cm]{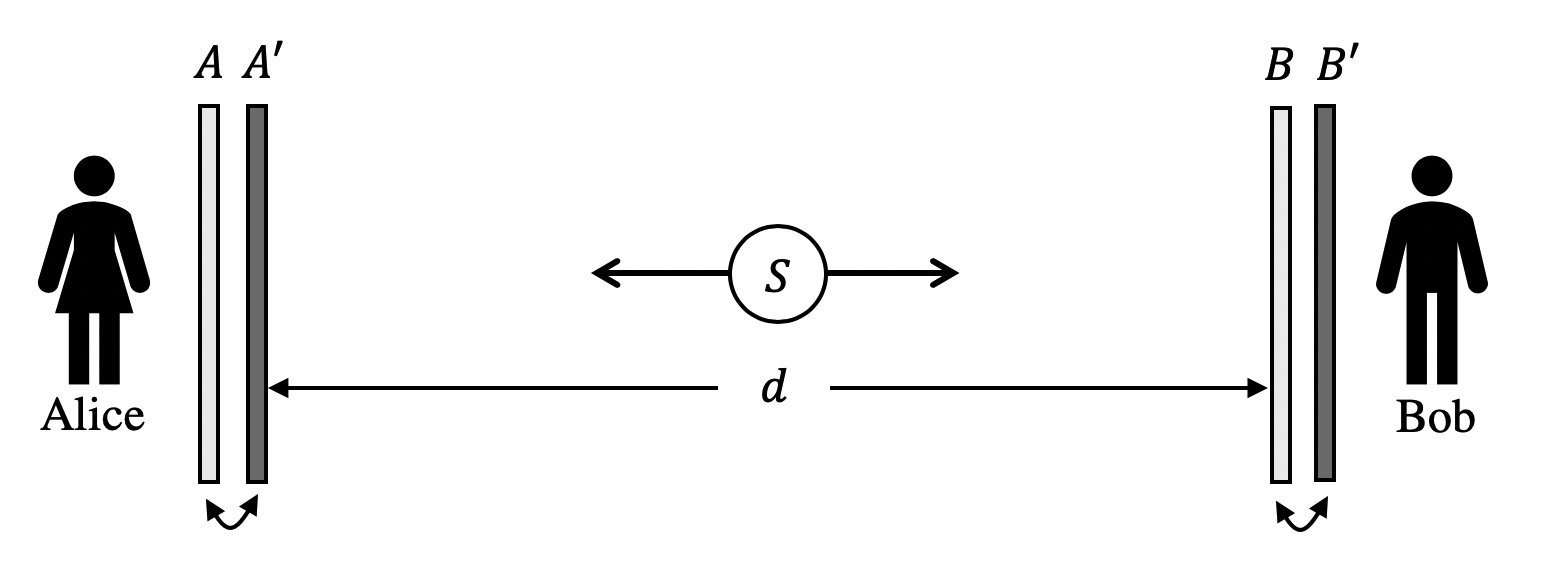} 
\caption{A schematic representation of the four coincidence measurements 
$AB$, $AB'$, $A'B$ and $A'B'$ performed by Alice and Bob. The source emits a particle that propagates either towards Alice or Bob, or towards both, in case it would be a quantum entity prepared in a superposition state. 
\label{Figure2}} 
\end{figure}

Consider a particle emitted by a source, it can either be a classical corpuscle or a quantum entity, like a photon or an electron, our analysis will equally hold for both situations. At some distance from the source, in opposite directions, Alice and Bob have placed their equipment to perform measurements. Alice's measurement $A$ consists in placing a detector screen and checking whether the particle is detected by it. If this is the case, the outcome is $A_1$, otherwise the outcome is $A_2$. Measurement $B$, performed by Bob, is the same, but carried out at his location. More specifically, following the emission of the particle, say at time $t_0$, Alice and Bob start their measurements at a subsequent predetermined time $t_1>t_0$, by attentively observing their screens up to time $t_2>t_1$, i.e., during a time interval 
$\Delta T = t_2-t_1$. In other words, a detection or non-detection outcome means that a trace was left or not by the particle, during the interval $\Delta T$. Measurements $A'$ and $B'$ are defined in the same way, with the only difference that Alice and Bob use for them a different kind of detection screen, which we assume has a greater detection efficiency or, equivalently, has a lower reflective power compared to the screen used in measurements $A$ and $B$. Again, outcomes $A'_1$ and $B'_1$ correspond to the situation of a detection within $\Delta T$, and $A'_2$ and $B'_2$ to the situation of a non-detection within that same time interval (see Figure~\ref{Figure2} for a schematic representation of the situation). 

Since only a single particle at a time is emitted by the source, Alice and Bob cannot both detect it during the time interval $\Delta T$, so we have: 
\begin{equation}
{\cal P}(A_1,B_1) ={\cal P}(A'_1,B_1)={\cal P}(A_1,B'_1)={\cal P}(A'_1,B'_1)=0.
\label{zero-probability}
\end{equation}
In case of a classical particle, we can assume that it is either emitted by the source towards Alice, say with probability ${1\over 2}$, or towards Bob, with same probability. If the particle is quantum, we can assume either the same, or that the particle is emitted in a symmetric (with respect to the source location) superposition state, formed by two wave-packets propagating in opposite directions: 
\begin{equation}
|\psi\rangle = {1\over \sqrt{2}}(|\psi^\leftarrow\rangle + |\psi^\rightarrow \rangle).
\label{initialstate}
\end{equation}
When Alice and Bob jointly perform their measurements, the following four combinations of screens are possible: $AB$, $AB'$, $A'B$ and $A'B'$. Because of the symmetry of the experiment, we then necessarily have: 
\begin{equation}
{\cal P}(A'_2,B_2) = {\cal P}(A_2,B'_2), \quad {\cal P}(A'_1,B_2)={\cal P}(A_2,B'_1), \quad {\cal P}(A'_2,B_1)={\cal P}(A_1,B'_2).
\label{symmetries}
\end{equation}
To simplify the notation, we set $p\equiv {\cal P}(A_2,B_2)$, $p'\equiv {\cal P}(A'_2,B_2) = {\cal P}(A_2,B'_2)$ and $p''\equiv {\cal P}(A'_2,B'_2)$. 
In other words, $p$ (respectively, $p'$, $p''$) is the probability of not detecting the particle during $\Delta T$, by either screen, when the screen combination is $AB$ (respectively, $A'B$ or $AB'$, $A'B'$). In view of (\ref{zero-probability}), we also have: 
\begin{eqnarray}
&&{\cal P}(A'_1,B_2)+{\cal P}(A'_2,B_1) = 1-p',\quad {\cal P}(A_1,B'_2)+{\cal P}(A_2,B'_1) = 1-p'\nonumber\\
&& {\cal P}(A_1,B_2)+{\cal P}(A_2,B_1) = 1-p,\quad {\cal P}(A'_1,B'_2)+{\cal P}(A'_2,B'_1) = 1-p''.
\label{xxx}
\end{eqnarray}
For the expectation values, we thus obtain: 
\begin{eqnarray}
&&E(A,B)={\cal P}(A_1,B_1) + {\cal P}(A_2,B_2) - {\cal P}(A_1,B_2)- {\cal P}(A_2,B_1)= 2p -1\nonumber\\
&&E(A,B')={\cal P}(A_1,B'_1) + {\cal P}(A_2,B'_2) - {\cal P}(A_1,B'_2)- {\cal P}(A_2,B'_1)= 2p' -1\nonumber\\
&&E(A',B)={\cal P}(A'_1,B_1) + {\cal P}(A'_2,B_2) - {\cal P}(A'_1,B_2)- {\cal P}(A'_2,B_1)=2p' -1\nonumber\\
&&E(A',B')={\cal P}(A'_1,B'_1) + {\cal P}(A'_2,B'_2) - {\cal P}(A'_1,B'_2)- {\cal P}(A'_2,B'_1)= 2p'' -1,
\label{expectations}
\end{eqnarray}
and for the quantity in the Bell-CHSH inequality, we find: 
\begin{eqnarray}
{\rm CHSH} &=& E(A',B')+E(A,B') + E(A',B) - E(A,B)\nonumber\\
&=& (2p'' -1)+2(2p' -1) - (2p -1) \nonumber\\
&=& 2(2p'+p''-p-1). 
\label{CHSH22}
\end{eqnarray}
Clearly, we will have a violation of the Bell-CHSH inequality, that is, a violation of $|{\rm CHSH}|\leq 2$, if $|1+(p-p'')-2p'|>1$. This will be the case if either $p-p''-2p'>0$ or $p-p''-2p'<-2$. The latter inequality implies that $p-p''<-2(1-p')<0$, which cannot hold, since we have assumed that the screens used in the $A$ and $B$ measurements are less efficient than those used in the $A'$ and $B'$ measurements, which means that it is during the joint measurement $AB$ that the particle has the highest probability of not being detected by the two screens, during the time interval $\Delta T$. On the other hand, the situation where the probability of not being detected (i.e., not being absorbed) is the lowest, is that of the joint measurement $A'B'$ (and of course for the two measurements $A'B$ and $AB'$ we are in an intermediate situation). Therefore, we will generally have $p>p''$. So, for violating the Bell-CHSH inequality, we remain with the condition: 
\begin{equation}
2p'<p-p''.
\label{conditionviolation}
\end{equation}

Here we immediately see that if, say, $p'$ is given by the arithmetic mean of $p$ and $p''$, i.e., $2p'=p+p''$, since $p+p''>p-p''$, there will be no violation. On the other hand, if $p'$ is given by the geometric mean $p'=\sqrt{pp''}$, then (\ref{conditionviolation}) becomes: 
\begin{equation}
2< \sqrt{p\over p''}- \sqrt{p''\over p} 
\label{conditionviolation2}
\end{equation}
and since $p$ and $p''$ can be varied one independently of the other (with the only constraint that $p>p''$), (\ref{conditionviolation2}) will be clearly obeyed in the limit $p''\to 0$, for $p>0$, hence the Bell-CHSH inequality will be violated. Note that in this regime, we also have $p'=\sqrt{pp''}\to 0$, so it follows from (\ref{CHSH22}) that $|{\rm CHSH}|\to 2|1+p|$, so that the maximum algebraic violation of the Bell-CHSH inequality will be reached if additionally $p\to 1$. 

The reasonableness of the assumption that $p'$ is given by the geometric mean will be discussed in the explicit calculation of Sec.~\ref{classical}, for the case of a classical particle.\footnote{Note that in a situation where the event of not being detected by screen $A$ is assumed to be independent of the event of not being detected by screen $B$, and $p_A$ and $p_B$ are the probabilities of these two independent events, then we can write $p=p_Ap_B$, and similarly, with obvious notation, $p'=p_{A'}p_B=p_Ap_{B'}$ and $p''=p_{A'}p_{B'}$, hence we would immediately have: ${p'}^2= p_{A'}p_Bp_Ap_{B'}= p_Ap_Bp_{A'}p_{B'}=pp''$.}
In this section, we continue our analysis by also considering Alice's and Bob's marginal probabilities. We define ($i=1,2$): 
\begin{eqnarray}
&&{\cal P}_B(A_i) \equiv {\cal P}(A_i,B_1)+{\cal P}(A_i,B_2),\quad {\cal P}_{B'}(A_i) \equiv {\cal P}(A_i,B'_1)+{\cal P}(A_i,B'_2),\nonumber\\
&&{\cal P}_B(A'_i) \equiv {\cal P}(A'_i,B_1)+{\cal P}(A'_i,B_2),\quad {\cal P}_{B'}(A'_i)\equiv {\cal P}(A'_i,B'_1)+{\cal P}(A'_i,B'_2),\nonumber\\
&&{\cal P}_A(B_i) \equiv {\cal P}(A_1,B_i)+{\cal P}(A_2,B_i), \quad {\cal P}_{A'}(B_i)\equiv{\cal P}(A'_1,B_i)+{\cal P}(A'_2,B_i), \nonumber\\
&&{\cal P}_A(B'_i) \equiv {\cal P}(A_1,B'_i)+{\cal P}(A_2,B'_i), \quad {\cal P}_{A'}(B'_i) \equiv{\cal P}(A'_1,B'_i)+{\cal P}(A'_2,B'_i).
\label{marginal probabilities}
\end{eqnarray}
The so-called marginal laws, express the requirement that the probabilities of Alice's outcomes do not depend on the measurements performed by Bob, and vice versa, that is ($i=1,2$): 
\begin{equation}
{\cal P}_B(A_i)={\cal P}_{B'}(A_i),\quad {\cal P}_B(A'_i)={\cal P}_{B'}(A'_i),\quad {\cal P}_A(B_i) ={\cal P}_{A'}(B_i),\quad {\cal P}_A(B'_i)={\cal P}_{A'}(B'_i).
\label{no-signaling}
\end{equation}
Considering that: 
\begin{eqnarray}
&&{\cal P}_B(A_1) = {\cal P}(A_1,B_2),\ {\cal P}_{B'}(A_1) = {\cal P}(A_1,B'_2),\ {\cal P}_A(B_1) ={\cal P}(A_2,B_1),\ {\cal P}_{A'}(B_1)= {\cal P}(A'_2,B_1)\nonumber\\
&&{\cal P}_B(A'_1) = {\cal P}(A'_1,B_2),\ {\cal P}_{B'}(A'_1) = {\cal P}(A'_1,B'_2),\ {\cal P}_A(B'_1) ={\cal P}(A_2,B'_1),\ {\cal P}_{A'}(B'_1)= {\cal P}(A'_2,B'_1)
\label{marginal selectivity-screens}
\end{eqnarray}
and in view of (\ref{xxx}), we obtain:
\begin{eqnarray}
&&[{\cal P}_B(A_1)-{\cal P}_{B'}(A_1)]+[{\cal P}_{B'}(A'_1)-{\cal P}_{B}(A'_1)]+[{\cal P}_A(B_1) -{\cal P}_{A'}(B_1)]+[{\cal P}_{A'}(B'_1)-{\cal P}_{A}(B'_1)]\nonumber\\
&&\quad =2p'-(p+p'').
\label{marginal selectivity-screens2}
\end{eqnarray}
In other words, the necessary condition for compliance with (\ref{no-signaling}) is that $p'$ is given by the arithmetic mean $p'={1\over 2}(p+p'')$. 
Note that if we insert the latter in (\ref{CHSH22}), we obtain $|{\rm CHSH}|= 2|2p''-1|\leq 2$, so there is no violation of the Bell-CHSH inequality, which means that the violation of the latter and of the marginal laws are intimately connected in our model.

\section{Classical entity}
\label{classical}

Let us assume that the entity emitted by the source is a classical point-like particle traveling towards Alice with velocity $v$. To reach Alice's screen, who is performing, say, measurement $A'$, it has to travel for a time ${1\over 2}{d\over v}$. Then, it can either be detected (i.e., absorbed) by Alice's screen, with probability $D'$, or reflected back by it, with probability $R'$, with $D' + R'=1$. In the latter situation, to reach Bob's screen, who we assume is performing measurement $B$, it has to travel for an additional time ${d\over v}$. Then, when interacting with Bob's screen, it can either be detected, with probability $D$, or reflected back towards Alice's screen, with probability $R$, with $D + R=1$. If it is reflected back, to reach once more Alice's screen, it has to travel for a further time ${d\over v}$, and the story possibly repeats, until either Alice or Bob, or none of them, will detect or not detect the particle within the observational time interval $\Delta T = t_2-t_1$. If we take $t_1={1\over 2}{d\over v}$ and $t_2=t_1 + {(2n-1)d\over v}$, then the particle can interact with Alice's screen at maximum $n$ times, with $n\geq 1$, 
because of the possible multiple reflections arising between the two screens, during the time interval $\Delta T$. Therefore, we can write:
\begin{eqnarray}
{\cal P}_{\leftarrow}(A'_1,B_2) &=& D' + R'RD' + R'RR'RD' + \dots + R'RR'R\cdots R'R D' \nonumber\\
&=& [1 + R'R + (R'R)^2 + \dots + (R'R)^{n-1}]D'= {1-(R'R)^n\over 1-R'R}(1-R').
\label{series-left}
\end{eqnarray}

We can reason in a similar way for the particle initially traveling towards Bob. In this case, it will reach Bob's screen after a flight time ${1\over 2}{d\over v}$, then be possibly reflected by it with probability $R$ and reach Alice's screen after an additional time ${d\over v}$. Reasoning as above, during the observational time interval 
$\Delta T$ the particle can interact with Alice's screen a maximum of $n$ times, $n\geq 1$, so Alice's detection probability is in this case: 
\begin{eqnarray}
{\cal P}_{\rightarrow}(A'_1,B_2) &=& RD' + RR'RD' + RR'RR'RD' +\dots + RR'RR'R\cdots R'RD' \nonumber\\
&=& R [1 + R'R + (R'R)^2 + \dots + (R'R)^{n-1}] D'= {1-(R'R)^n\over 1-R'R}R(1-R')
\label{Alice-detection-A'B-right}
\end{eqnarray}
Bringing the above two probabilities together, we obtain: 
\begin{equation}
{\cal P}(A'_1,B_2) = {1\over 2} {\cal P}_{\leftarrow}(A'_1,B_2) + {1\over 2}{\cal P}_{\rightarrow}(A'_1,B_2) = {1\over 2} {1-(R'R)^n\over 1-R'R}(1 + R)(1-R').
\label{Alice-detection-A'B}
\end{equation}
Reasoning in a specular way, we find for the probability that the particle is detected by Bob, and therefore not by Alice:
\begin{equation}
{\cal P}(A'_2,B_1) = {1\over 2} {1-(R'R)^n\over 1-R'R} (1 + R')(1-R).
\label{Bob-detection-A'B}
\end{equation}
Adding these two probabilities, we obtain:
\begin{eqnarray}
{\cal P}(A'_1,B_2)+{\cal P}(A'_2,B_1)&=&{1\over 2} {1-(R'R)^n\over 1-R'R} [(1 + R)(1-R')+(1 +R')(1-R)]\nonumber\\
&=&{1\over 2}{1-(R'R)^n\over 1-R'R} (2 -2R'R) = 1-(R'R)^n.
\label{sum-of-prob}
\end{eqnarray}
Considering that ${\cal P}(A'_1,B_1)=0$, we deduce that: 
\begin{equation}
{\cal P}(A'_2,B_2) = 1- [{\cal P}(A'_1,B_2)+{\cal P}(A'_2,B_1)+{\cal P}(A'_1,B_1)]=(R'R)^n\equiv p'.
\label{prob'}
\end{equation}
For symmetry reasons, we find the same result for ${\cal P}(A_2,B'_2)$, and for measurements $AB$ and $A'B'$ we have: ${\cal P}(A_2,B_2) = R^{2n}\equiv p$ and ${\cal P}(A'_2,B'_2) = {R'}^{2n}\equiv p''$, so that $p'$ is given by the geometric mean: 
\begin{equation}
p'= (R'R)^n =\sqrt{{R'}^{2n}{R}^{2n}} =\sqrt{p'' p}.
\label{gm}
\end{equation}
Hence, following the discussion of the previous section, marginal selectivity and the Bell-CHSH inequality will be violated and the latter can also be maximally violated, in the limits $R\to 1$ and $R'\to 0$

\section{Quantum entity}
\label{quantum}

If the entity emitted by the source is quantum, one has to reason in terms of wave-packets. Additional effects are therefore also possibly involved in this case, like resonance effects resulting from the fact that forward and backward wave-packets can superpose and interfere in the regions of the detection screens. This is reflected in the fact that when considering the different paths that the entity can follow, when multiply reflected back and forth between Alice's and Bob's screens, one does not have to add probabilities but (complex valued) probability amplitudes. However, this will not alter qualitatively our analysis above, but only introduce a modulation of the possible values for the non-detection probabilities $p$, $p'$, and $p''$, depending on possible resonant conditions that can be obtained. 

More precisely, if ${\cal D'}$ denotes the probability amplitude for the entity to be detected by Alice's screen, assuming she is performing measurement $A'$, and ${\cal R'}$ the probability amplitude for the entity not to be detected but reflected to the right by Alice's screen and then reach Bob's screen, we have $|{\cal D'}|^2+|{\cal R'}|^2=1$. Similarly, assuming Bob is performing measurement $B$, we can define the probabilities amplitudes ${\cal D}$ and ${\cal R}$ of being detected by his screen or being reflected back to the left and reach again Alice's screen, respectively, with $|{\cal D}|^2+|{\cal R}|^2=1$. 

Then, a similar calculation as in (\ref{series-left}) can be done to determine the probability amplitude ${\cal A}_{\leftarrow}(A'_1,B_2)$, for Alice detecting the entity and Bob not detecting the entity, when the latter is emitted by the source towards Alice. Assuming that the wave-packet is well peaked at about energy $E$, with $E={\hbar^2k^2\over 2m}$, and taking here for simplicity an observational time-interval $\Delta T={3d\over v}$, with $v={\hbar k\over m}$, we will typically have the two contributions ${\cal A}_{\leftarrow}(A'_1,B_2)\approx {\cal D'} + {\cal D'}{\cal R}{\cal R'}$, so that ${\cal P}(A'_1,B_2)\approx |{\cal D'}|^2|1 + {\cal R}{\cal R'}|^2$. Similarly, for a wave-packet initially propagating towards Bob, we have: 
${\cal A}_{\rightarrow}(A'_1,B_2)\approx {\cal R}{\cal D'}$, hence: ${\cal P}_{\rightarrow}(A'_1,B_2)\approx |{\cal D'}|^2|{\cal R}|^2$. So, 
\begin{equation}
{\cal P}(A'_1,B_2) = {1\over 2} {\cal P}_{\leftarrow}(A'_1,B_2) + {1\over 2}{\cal P}_{\rightarrow}(A'_1,B_2) \approx {1\over 2} |{\cal D'}|^2(|1 + {\cal R}{\cal R'}|^2 +|{\cal R}|^2).
\label{Alice-detection-A'Bquantum}
\end{equation}
Reasoning in a similar way, we find ${\cal P}(A'_2,B_1) \approx {1\over 2} |{\cal D}|^2(|1 + {\cal R}{\cal R'}|^2 +|{\cal R'}|^2)$. We thus obtain: 
\begin{equation}
p' = 1- [{\cal P}(A'_1,B_2)+{\cal P}(A'_2,B_1)]\approx 1-{1\over 2} |{\cal D'}|^2(|1 + {\cal R}{\cal R'}|^2 +|{\cal R}|^2) - {1\over 2} |{\cal D}|^2(|1 + {\cal R}{\cal R'}|^2 +|{\cal R'}|^2).
\label{prob'-q}
\end{equation}
Clearly, we also have: 
\begin{equation}
p \approx 1- |{\cal D}|^2(|1 + {\cal R}^2|^2 +|{\cal R}|^2), \quad p'' \approx 1- |{\cal D'}|^2(|1 + {\cal R'}^2|^2 +|{\cal R'}|^2).
\label{prob-p-p''-q}
\end{equation}

Considering the limit $|{\cal D'}|\to 1$ (and consequently $|{\cal R'}|\to 0$), (\ref{prob'-q}) gives $p'\to 1-{1\over 2}(1+|{\cal R}|^2)- {1\over 2} |{\cal D}|^2=0$ and (\ref{prob-p-p''-q}) also gives $p''\to 0$. Hence (\ref{conditionviolation}) is obeyed in this limit, and for $|{\cal D}|\to 0$, we also have $p\to 1$ and the Bell-CHSH inequality is maximally violated. 

The situation is the same (within the limits of our semiclassical discussion) if the source emits a wave-packet in a superposition state (\ref{initialstate}), with the two packets propagating with opposite (group) velocities. Indeed, since they will reach Alice's (respectively, Bob's) screen in different times, when considering the outcome $(A'_1,B_2)$ (respectively, $(A'_2,B_1)$), they can be treated as non-interfering alternatives.

More detailed quantum calculations can of course be worked out, which are however beyond the scope of the present article. For instance, the screens' absorbing powers can be modeled by adding a dissipative interaction term to the Hamiltonian (according to so-called optical model \cite{Feshbach1958}). More precisely, if $H$ is the Hamiltonian governing the evolution of the quantum entity in the presence of two totally reflective screens, the joint measurement $A'B$ can be described by adding a dissipative interaction term $-i\lambda' P_{\rm Alice} -i\lambda P_{\rm Bob}$, where $P_{\rm Alice} $ and $P_{\rm Bob} $ are the projection operators onto the set of states localized in the region where Alice's and Bob's screen are located, respectively. The two positive constants $\lambda'$ and $\lambda$ then determine the absorbing powers of Alice's and Bob's screens, and $p'= {\cal P}(A'_2,B_2)={\cal P}_{t_2}$, with ${\cal P}_{t_2}=\langle\psi_{t_2} |\psi_{t_2}\rangle$, $|\psi_{t_2}\rangle = e^{-\frac{i}{\hbar}H(\lambda',\lambda)(t_2-t_0)}|\psi_{t_0}\rangle$, $H(\lambda',\lambda)=H -i\lambda' P_{\rm Alice} -i\lambda P_{\rm Bob}$. In other words, by studying ${\cal P}_{t}$, which obeys ${d {\cal P}_t\over dt}=-{2\over \hbar} \langle\psi_t |( \lambda' P_{\rm Alice} +\lambda P_{\rm Bob})|\psi_t\rangle$, with ${\cal P}_{t_0}=1$, one can estimate $p'$, and similarly for the other three joint measurements.

\section{Signaling}

As known, when considering a quantum two-entity entangled system, the four joint measurements $AB$, $A'B$, $AB'$ and $A'B'$ are usually described as tensor product observables: $A\otimes B$, $A'\otimes B$, $A\otimes B'$ and $A'\otimes B'$. This automatically guarantees that the marginal laws are obeyed, and this is the reason why it is usually considered that quantum mechanics cannot contradict relativity, even if the collapse of the wavefunction is an instantaneous phenomenon. The typical reasoning goes as follows. One first assumes that Alice and Bob, who are located at great distance from one another, share a sufficiently large number of identically prepared entangled states, where by `sufficiently large' we mean `large enough to be able to produce a robust statistics of outcomes, when performing multiple times the same joint measurement, so as to allow to unambiguously distinguish it from other possible joint measurements'. If the statistics that Alice obtains depend on the measurement Bob is jointly performing at his location, Bob has then in principle a way to use this fact to communicate with Alice \cite{Ballentine1987}. 

One can think for instance that Alice and Bob have arranged things in such a way that they can perform all these identical measurements, providing the necessary statistics, in a parallel way, so that the execution time of a single coincidence measurement can be assumed to approximately correspond also to the execution time of performing a large number of them. If one adds to the above the hypothesis that Alice and Bob can actualize at the same time their outcomes (in the reference frame in which their apparatuses are at rest), that is, that the collapse of the entangled state would be instantaneous and independent of the distance separating them, one can easily imagine how a superluminal form of communication could take place, thanks to the violation of the marginal laws (\ref{no-signaling}), which are therefore also called no-signaling conditions. 

Our model flagrantly violates the marginal laws, but of course cannot be used to achieve a superluminal communication. The observational time interval being $\Delta T={(2n-1)d\over v}$, its minimal value, for $n=1$, is ${d\over v}$, hence no communication can here become superluminal, even when $v\to c$. Of course, one might also consider a shorter observational interval $\Delta T< {d\over v}$. But then, if the classical particle is emitted by the source towards Alice, it can interact with her screen only once, and even if reflected back it will not have the time to interact with Bob's screen. Similarly, if the particle is emitted by the source towards Bob, it can interact with his screen only once, and even if reflected back it will not have the time to interact with Alice's screen. Therefore, we have in this case: ${\cal P}_{\leftarrow}(A'_1,B_2)= D'$ and ${\cal P}_{\rightarrow}(A'_1,B_2)=0$, hence ${\cal P}(A'_1,B_2)={1\over 2}{\cal P}_{\leftarrow}(A'_1,B_2)+{1\over 2}{\cal P}_{\rightarrow}(A'_1,B_2)= {1\over 2}D'$, and similarly, ${\cal P}(A'_2,B_1)={1\over 2}D$, so that, according to (\ref{prob'}), $p'=1-{1\over 2}(D+D')= {1\over 2}(R'+R)$. In the same way, we find $p=R$ and $p''=R'$. This means that instead of a geometric mean, we now have an artithmetic mean, $p'={1\over 2}(p+p'')$, hence, there will be no violation of the marginal laws and Bell-CHSH inequality if $\Delta T< {d\over v}$. And the same is true for a quantum entity, at least in the ambit of a semiclassical reasoning. 

Of course, in the full quantum treatment, the spreading of the wave-packets will also play a role. Indeed, as is well-known \cite{Jaworsky1989}, the probability distribution of a quantum entity, initially localized in a finite region at time $t_0$, will immediately evolve by spreading in space, in such a way that there will be a non-zero probability of finding it in any other region of space for (almost) all times $t>t_0$. But the relevance of this additional effect, produced by the wave-packet spreading, is unlikely to be significant enough from a statistical viewpoint, as what primarily matters in terms of detection time by the screens is the group velocity at which the wave-packets' peaks move.

\section{Conclusion}

To conclude, we have presented a detection model where the measurements' outcomes are about detecting or not detecting a particle, which can either be a classical corpuscle or a quantum micro-entity. Our idealized experimental situation exploits to its advantage the so-called `detection loophole' to produce correlations that are able to violate the Bell-CHSH inequality. In other words, the non-ideal intrinsic efficiency of the detectors is viewed here not as a limitation, but as a resource, in the sense that a non-detection (during the predetermined observational time interval) is interpreted as a bona fide outcome, and not as a measurement without an outcome. Finally, we have shown that the no-signaling conditions (marginal laws) are also strongly and intrinsically violated in our model, and therefore can generally be violated in quantum mechanics, without this being necessarily in conflict with the relativistic principles \cite{AertsEtAl2019, Sassoli2019}.

We emphasize that one should not think of the experimental situation described in this article as a model where Alice and Bob would merely be allowed to communicate in order to violate the Bell-CHSH inequality. It is well known that Bell's correlations can be simulated by allowing some form of communication between Alice and Bob (see \cite{Toner2003} and the references cited therein). In our situation there is no explicit and voluntary communication occurring between Alice and Bob. Simply, Alice's choice of detector screen affects Bob's detection probabilities, and vice versa, thus producing the violation of the no-signaling conditions. This is so because of the nature of the performed measurements and, more importantly, of the fact that they are conjunctly performed on a same single entity. In other words, this is not just a model where some slower-than-light signaling is introduced by hand to obtain a violation of the Bell-CHSH inequality (with Alice producing an output, sending some information to Bob, who would then use it to subsequently produce his correlated output, or vice versa) but a paradigmatic situation where correlations are created by jointly and simultaneously operating on a same undivided entity.

Our analysis indicates that a single-entity situation already contains those 
core aspects that are responsible for producing a violation of the Bell-CHSH inequality in bipartite systems, as well as a possible violation of the no-signaling conditions, as observed in many experiments \cite{AdenierKhrennikov2007,DeRaedt2012,DeRaedt2013,AdenierKhrennikov2016,Bednorz2017,Kupczynski2017}. Of course, when studying entanglement in bipartite systems, additional elements come into play in the way the phenomenon is usually interpreted, in particular because of our tendency in viewing spatially separated entities as not anymore forming a single whole. On the other hand, our model (and our additional analysis in \cite{AertsEtAl2019}) indicates that two entangled entities can be understood as forming an undivided and interconnected whole, from which (when joint measurements are performed on it) correlations can be created, in a way that both the Bell-CHSH inequality and the marginal laws can be violated. 

From the viewpoint of the quantum formalism, this means that it is not possible anymore to represent the joint measurements by using a unique tensor product decomposition, and we refer the reader to \cite{AertsEtAl2019,as2014} for a more general Hilbertian representation. As regards the nature of quantum entanglement, it follows that non-locality can and probably should be understood as the consequence of the non-spatial nature of entangled systems, which in turn would explain why micro-entities, when entangled, are able to form an effective single-entity, despite of the fact that they can be separated by arbitrarily large spatial distances. This should not be surprising, since even when we experiment with single-entities, like in the remarkable neutron interferometry experiments performed in the mid-seventies of the last century \cite{Rauch1975,Werner1975}, and data are attentively analyzed \cite{Aerts1999,Sassoli2017}, it is quite obvious that they do not behave as particles, waves or fields, but as genuine non-spatial elements of our physical reality (see also the perspective offered by Kastner's possibilistic transactional interpretation of quantum mechanics \cite{k2013}). It is this same non-spatiality, manifesting itself already at the single-entity level, which in our view can explain why bipartite systems are able to form interconnected unities that can violate not only the Bell-CHSH inequality but also possibly the marginal laws.
\\

\noindent {\bf Acknowledgements}\ This work was partially supported by QUARTZ (Quantum Information Access and Retrieval Theory), the Marie Sklodowska-Curie Innovative Training Network 721321 of the European Union’s Horizon 2020 research and innovation programme.

\end{document}